\documentclass[referee]{aa} 
\usepackage{graphicx}
\usepackage{txfonts}
\newcommand{\Lsol}{L$_{\odot}$}
\newcommand{\Msol}{M$_{\odot}$}
\newcommand{\Msold}{M$_{\odot}$\,yr$^{-1}$}
\newcommand{\Vlsr}{V$_{\rm lsr}$}
\newcommand{\Vexp}{V$_{\rm exp}$}
\newcommand{\Teff}{T$_{\rm eff}$}
\newcommand{\Rph}{R$_{\rm ph}$}

\newcommand{\kms}{km\,s$^{-1}$}
\newcommand{\HI}{H\,{\sc {i}}~}
\newcommand{\MHI}{M$_{\rm H \sc I}$}

\newcommand{\lsim}{\rlap{$<$}{\lower 1.0ex\hbox{$\sim$}}}
\newcommand{\gsim}{\rlap{$>$}{\lower 1.0ex\hbox{$\sim$}}}
\newcommand{\ub}{\underbar}
\begin{document}
 \title{Circumstellar \HI and CO around the carbon stars\\
V1942\,Sgr and V\,CrB}

   \author{Y. Libert\inst{1},  
	   E. G\'erard\inst{2},
	   C. Thum\inst{3},
	   J.M. Winters\inst{3},
           L.D. Matthews\inst{4},
	   \and T. Le\,Bertre\inst{1}
          }

   \institute{LERMA, UMR 8112, Observatoire de Paris, 
	      61 Av. de l'Observatoire, 75014 Paris, France 
         \and
	      GEPI, UMR 8111, Observatoire de Paris, 
	      5 Place J. Janssen, 92195 Meudon Cedex, France
         \and
              IRAM, 300 rue de la Piscine, 
	      38406 St. Martin d'H\`eres, France
         \and
              MIT Haystack Observatory, Off Route 40,
              Westford, MA 01886, USA
             }

   \date{Received 11 September 2009/ Accepted 20 October 2009}

   \titlerunning{\HI and CO around carbon stars}
   \authorrunning{Libert et al.}

 
  \abstract
{The majority of stars that leave the main sequence are undergoing extensive 
mass loss, in particular during the asymptotic giant branch (AGB) 
phase of evolution. Observations show that the rate at which this phenomenon 
develops differs highly from source to source, so that the time-integrated 
mass loss as a function of the initial conditions (mass, metallicity, etc.) 
and of the stage of evolution is presently not well understood. }
{We are investigating 
the mass loss history of AGB stars by observing the molecular and atomic 
emissions of their circumstellar envelopes.} 
{In this work we have selected two stars that are on the thermally pulsing 
phase of the AGB (TP-AGB) and for which high quality data in the CO rotation 
lines and in the atomic hydrogen line at 21 cm could be obtained.} 
{V1942 Sgr, a carbon star of the Irregular variability type, shows a complex 
CO line profile that may originate from a long-lived wind at a rate of 
$\sim$ 10$^{-7}$ \Msold, and from a young (\lsim10$^4$\,years) fast outflow 
at a rate of $\sim$ 5 10$^{-7}$ \Msold. Intense \HI emission indicates 
a detached shell with 0.044 \Msol ~of hydrogen. This shell probably results 
from the slowing-down, by surrounding matter, of the same long-lived wind 
observed in CO that has been active during  $\sim$\,6\,10$^{5}$ years.
On the other hand, the carbon Mira V CrB is 
presently undergoing mass loss at a rate of 
2\,10$^{-7}$\,\Msold, but was not detected in H\,{\sc {i}}. The wind is mostly 
molecular, and was active for at most 3\,10$^{4}$ years, with an integrated 
mass loss of at most 6.5\,10$^{-3}$\,\Msol.}
{Although both sources are carbon stars on the TP-AGB, they appear to develop 
mass loss under very different conditions, and a high rate of mass loss 
may not imply a high integrated mass loss.}


   \maketitle
%

\section{Introduction}

Low- to intermediate-mass stars, at the end of their main-sequence evolution, 
become first hydrogen shell-burning red giants (RGB 
$-$Red Giant Branch$-$ 
stars), then hydrogen and helium shell-burning red giants (AGB 
$-$Asymptotic Giant Branch$-$ stars). 
In this second phase they may undergo mass loss 
at a very large rate ($>$ 10$^{-8}$ \Msold), even so large that 
it has a decisive effect on their late evolution (Olofsson 1999). 
They are surrounded by expanding envelopes  
of gas and dust that have been extensively observed with radio 
molecular lines and infrared continuum emission. These tracers are used 
to estimate mass-loss rates. However the estimates are somewhat 
ambiguous because the mass-loss rate of a given source may vary on 
many different timescales. The mass change as a function 
of time due to mass loss is thus 
difficult to evaluate, and to connect with stellar evolution models. 
Furthermore molecular lines probe an extent of the circumstellar shell (CS)
that is limited by photo-dissociation, and therefore furnish information 
mainly on the inner parts of CSs, and on the recent mass loss. 

To try to circumvent these difficulties we have started a systematic programme 
of observations of red giants in the line of atomic hydrogen at 21 cm. 
We have published some of our results in several recent papers, and 
first reports on sizeable samples have been presented by G\'erard \& Le~Bertre 
(2006, hereafter GL2006) and Matthews \& Reid (2007, hereafter MR2007). 
A major difficulty of this programme is the confusion caused by the 
21 cm emission from the Interstellar Medium (ISM) that is located on the
same line-of-sight as the source of interest. This has a strong impact on the
observations which have to be conducted with a specific approach, and on the 
data processing that aims at providing spectra corrected for the ISM emission.
Perhaps more confounding, as circumstellar matter is expected to be 
at some stage injected in the ISM, the confusion by the {\ub {local}}
ISM might actually be at least partly of stellar origin, i.e. caused 
by material ejected at an earlier stage of evolution.  

In addition to observing systematically the \HI line at 21 cm in a large 
sample of sources with different properties, it is also important to choose 
objects for which the Galactic confusion is low and/or can be tracked easily, 
and therefore corrected accurately. 
The detailed study of such spectra should serve as a guide to 
exploit the data that are obtained in more difficult situations.  

Here we present our results on two carbon stars, V1942 Sgr and V CrB, for 
which the confusion is not a serious problem, and which have \HI properties 
that differ radically. Both are N-type carbon stars (CGCS 4229 and CCCS 2293, 
respectively) and have already been detected in CO rotational lines (Olofsson 
et al. 1993a). However the only available CO spectrum of V1942 Sgr had a poor 
signal-to-noise ratio, and for our study it appeared essential to also obtain 
new CO data of high quality.


\section{V1942\,Sgr}

V1942\,Sgr is classified as a long-period irregular variable (type Lb). 
Lebzelter \& Obbrugger (2009) have compared the lightcurve properties of 
Semi-Regular (SR) and Lb variables, and concluded that 
Lb stars can be seen as an extension of the SRs towards shorter periods 
and smaller amplitudes.
V1942 Sgr is a carbon star on the TP-AGB with a C/O ratio around 1.12 
(Olofsson et al. 1993b).  
Bergeat et al. (2001) estimate its effective temperature, \Teff, at 2960 K.
The parallax measured by Hipparcos (1.87 $\pm$\,0.51 mas, van Leeuwen 2007) 
places it at 535 pc, which implies a luminosity of 5200 \Lsol. 
From the data obtained by IRAS in the mid-infrared there is no direct evidence 
that the star is undergoing mass loss (the low resolution 8-22 $\mu$m 
spectrum is featureless).
However, Olofsson et al. (1993a) discovered emission in the CO(1-0) 
rotational line centered at \Vlsr = --31.5 \kms, close to the expected 
radial velocity of V1942 Sgr (\Vlsr = --32.0 \kms ~from the General Catalogue 
of Stellar Radial Velocities). 
The expansion velocity, \Vexp = 12.4 \kms, is surprisingly large 
for an Lb variable. From this spectrum Sch\"oier \& Olofsson (2001) 
derive a mass loss rate of $\sim$ 2.5 10$^{-7}$ \Msold ~(at 535 pc).
Extended emission associated with V1942\,Sgr was discovered 
by IRAS (Young et al. 1993a). The 60 $\mu$m data show a resolved shell 
of external radius 3.2$'$, i.e. 0.50 pc.

\subsection{\HI observations}\label{HI}

\begin{figure}
\centering
\includegraphics[width=8cm,angle=-90]{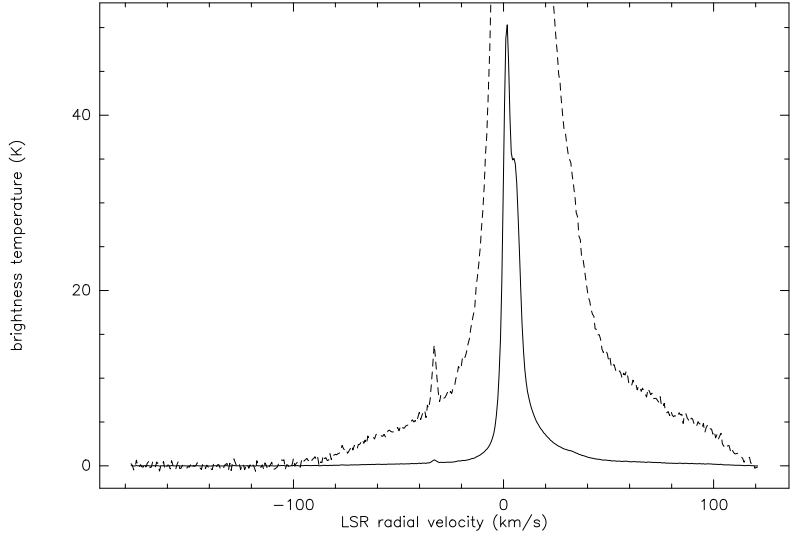}
\caption{Frequency-switched \HI 21 cm 
spectrum obtained with the NRT on the position 
of V1942\,Sgr. The spectrum enlarged by a factor 20 is also shown as a 
dashed line. The emission from V1942\,Sgr is clearly detected at --33\,\kms.}
\label{fswitchV1942}
\end{figure}

The \HI emission was observed with the Nan\c cay Radio Telescope (NRT), 
between March 2007 and July 2009, for a total of 85 hours. 
The NRT beamwidth (FWHM) at 21 cm is 4$'$ in right ascension (RA) and 22$'$  
in declination (Dec). An 'on' source frequency-switched spectrum is 
presented in Fig.~\ref{fswitchV1942}. The main emission peaks at 50 K around 
\Vlsr = 0 \kms. A narrow emission feature with a peak of $\sim$\,0.3\,K, 
centered at --33 \kms, is visible on top of the 0.4 K blue wing 
of the main peak near 0 \kms.

Position-switched spectra were also obtained with on-position taken at 
the position of V1942\,Sgr and off-positions, at $\pm$\,2$'$, $\pm$\,4$'$,  
$\pm$\,6$'$, $\pm$\,8$'$, $\pm$\,10$'$, $\pm$\,12$'$, $\pm$\,16$'$, 
$\pm$\,24$'$, and $\pm$\,32$'$.
The comparison between the spectra obtained for different values of the throw 
shows that the interstellar emission varies linearly with offset in the 
velocity range from --100 to --20 \kms. It means that the \HI 
background emission around V1942\,Sgr shows a gradient in RA which is 
constant for each velocity. This situation is similar to that 
encountered for EP Aqr and Y CVn (Le~Bertre \& G\'erard 2004, Figs.~3 and 7). 
In such a case the source emission can be extracted from the position-switched 
spectra by subtracting the contribution of the interstellar emission, 
which is estimated by interpolation between the two extreme off-positions. 

\begin{figure}
\centering
\includegraphics[width=8cm,angle=-90]{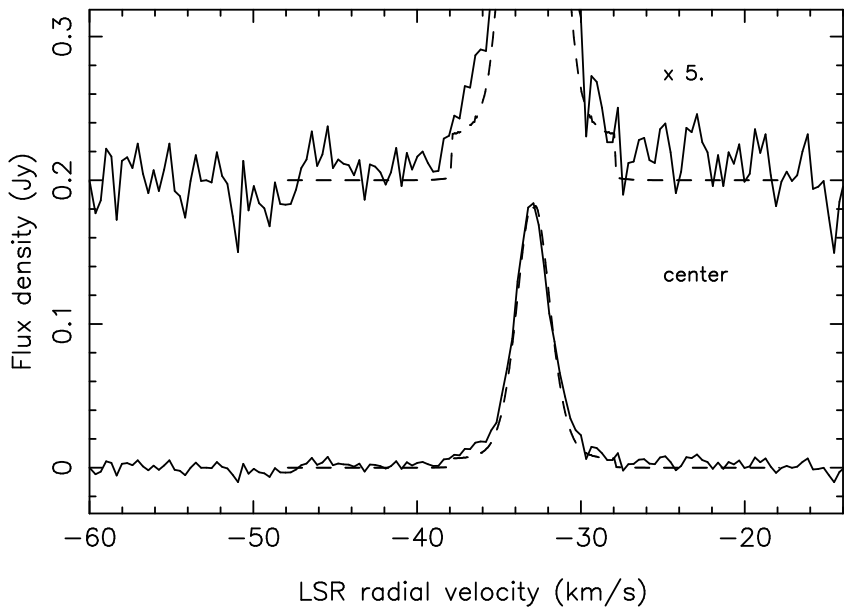}
\caption{\HI line profile of V1942\,Sgr. The spectrum has been smoothed 
to a resolution of 0.32 \kms. The dashed line is a fit obtained 
with the model described in Sect.~4.1.}
\label{centV1942}
\end{figure}

The intensity of the emission detected from the source in the 
position-switch spectra is constant with throw from $\pm$\,4$'$ 
to $\pm$\,32$'$. Therefore the source is mostly confined to the central beam 
(i.e. $\pm$\,2$'$ in RA; see GL2006, Sect. 2.1).  
The spectrum obtained at the star's position is shown in Fig.~\ref{centV1942}. 
It has a shape similar to that obtained on Y CVn (Libert et al. 2007) 
with a narrow emission line superposed 
on a pedestal extending from --39 to --27 \kms. The narrow emission is 
centered at \Vlsr\,=\,--32.9 \kms ~and has a quasi-gaussian profile of width 
2.95\,\kms ~(FWHM) and peak intensity 168 mJy. The pedestal is also 
centered at $\sim$ --33\,\kms ~and has an intensity of $\sim$\,6$\pm$2\,mJy. 
The peak intensity at the central position (178 mJy) agrees 
with that observed on the frequency-switched spectrum (cf. Fig.~1, with a 
conversion factor 2.15\,K/Jy for the NRT at 21 cm). We have also 
searched for \HI emission at blueshifted velocities down to --48 \kms, 
and redshifted velocities up to --18 \kms 
~(see the CO spectra in Sect.~2.2). We set an upper limit of 2 mJy for 
emission over this range. Nevertheless, we suspect residual features 
at --46, --26, and --24 \kms, possibly peaking at $\sim$ 3 mJy.

\begin{figure*}
\includegraphics[width=11.5cm,angle=-90]{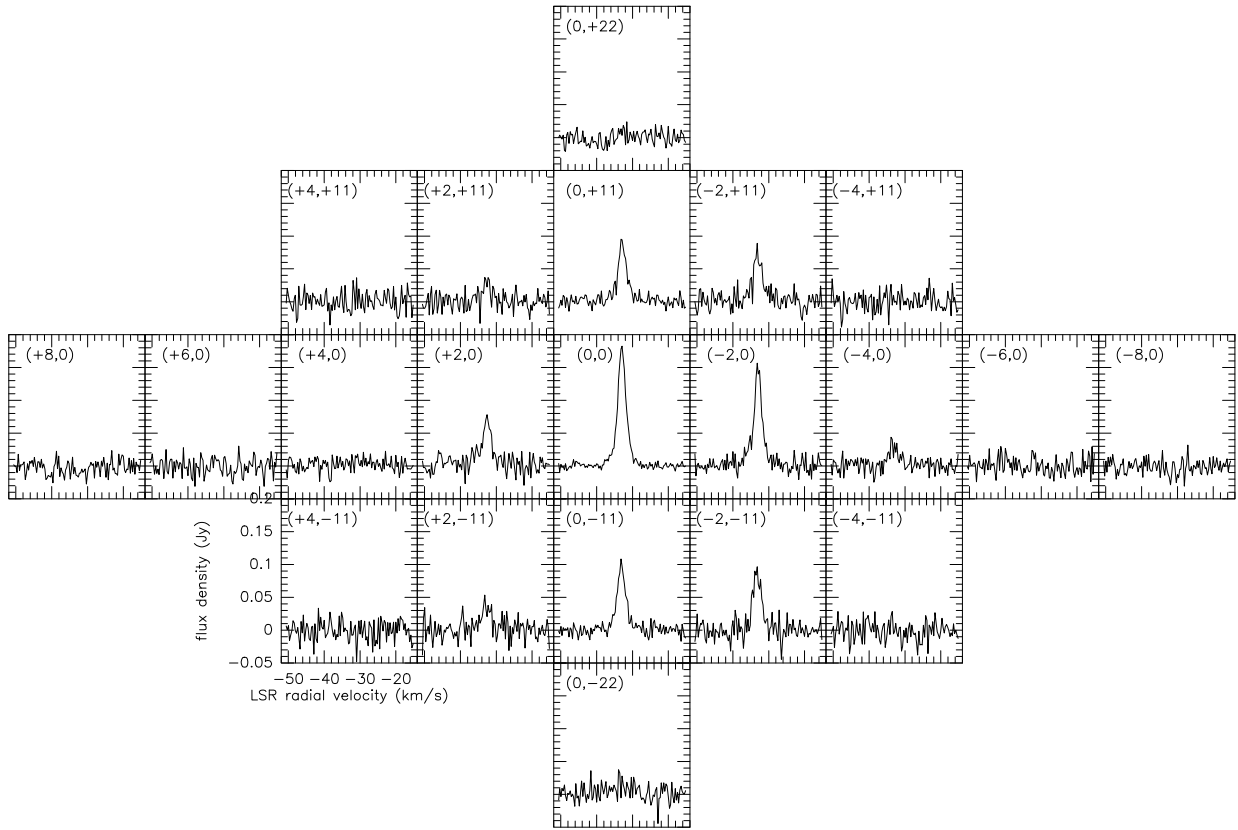}
\caption{Map of the 21 cm \HI emission of V1942\,Sgr. In each box the label at 
upper left gives the position (RA, Dec) with respect to the central star in 
arcminutes.}
\label{HImapV1942}
\end{figure*}

The spectra of the source in the off-positions are then determined 
by subtracting the individual position-switched spectra 
(on--off) and the contribution of the interstellar emission (assuming 
that it varies linearly with RA) from the central spectrum. 
Furthermore we have obtained data with the on-positions at 11$'$ north 
and 11$'$ south, and off-positions at $\pm$\,2$'$, $\pm$\,4$'$, 
and $\pm$\,12$'$, and with the on-positions  
at 22$'$ north and 22$'$ south, and off-positions at $\pm$\,8$'$. 
All these data are used to construct the flux density map of the source 
presented in Fig.~\ref{HImapV1942}. 

On this map we see that the intensity at --2$'$ west is almost the same 
as on the star, and therefore conclude that the source is slightly 
offset west from the stellar position. Assuming a gaussian distribution of 
the intensity, we estimate that the \HI source is centered at 0.6$'$ 
($\pm$\,0.1$'$) west in RA and at 0$'$ ($\pm$\,1$'$) in Dec. The size (FWHM) 
would then be $\sim$~1.3$'$ in RA and $<$5$'$ in Dec. The integrated flux 
in the map is 0.71\,Jy$\times$\kms. Assuming that the emission is 
optically thin and that atomic hydrogen is at a temperature well above 
the background ($\lsim$~0.4\,K + 4.2\,K, Reich \& Reich 1986), and using 
the standard relation, \MHI = $2.36 10^{-7} d^2 \int S_{\rm V} dV$, 
with \MHI ~in \Msol, $d$ in pc, and $\int S_{\rm V} dV$ ~in 
Jy$\times$\kms, this flux  
translates to 0.048 \Msol ~of atomic hydrogen at 535 pc. 

\subsection{CO observations}\label{CO}

CO observations of V1942\,Sgr were obtained at the IRAM 30-m telescope 
equipped with EMIR (Eight MIxer Receiver) on June 23, 2009 under average 
conditions (precipitable water vapor, pwv $\sim$\,10\,mm).
The two rotational lines, 1-0 and 2-1, were observed in parallel.
The four EMIR bands were selected to detect the two orthogonal polarizations 
at 115.2712 and 230.5380 GHz (T$_{\rm sys} \sim$\,400\,K, $\sim$\,800\,K, 
respectively). High spectral resolutions of 20 kHz 
and 40 kHz respectively (hence 0.05 \kms) were obtained with the 
VESPA (Versatile SPectrometer Array) backends.
The telescope beamwidths are 21$''$ (at 115 GHz) and 11$''$ (at 230 GHz), 
and the observations were made in the wobbler-switching mode using a throw 
of 60$''$ in azimuth.

\begin{figure}
\centering
\includegraphics[width=8.6cm,angle=0]{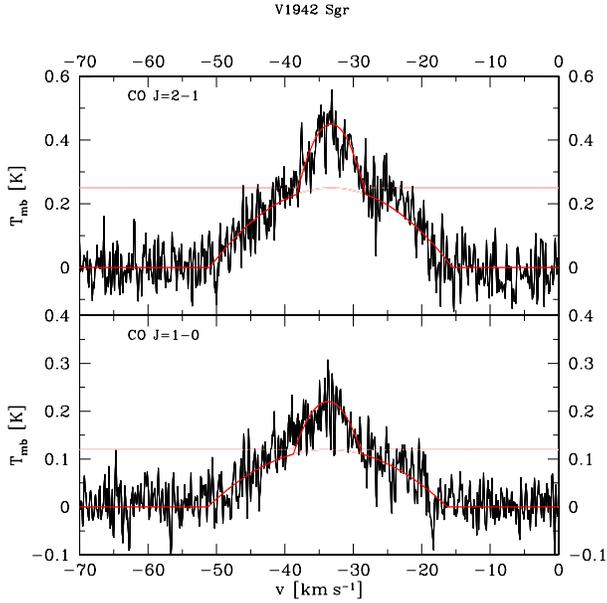}
\caption{CO (2-1, upper panel) and (1-0, lower panel) spectra of V1942\,Sgr 
obtained with the IRAM-30m telescope. 
The fits used to derive the wind parameters are also shown 
(see Table~\ref{COfit}).}
\label{COspectra}
\end{figure}

The spectra are shown in Fig.~\ref{COspectra}. These new spectra reveal 
that the line profiles are composite with two components centered on 
the same central velocity, but with different widths, like those observed 
by Knapp et al. (1998) and Winters et al. (2003) in several late-type giants, 
mostly oxygen-rich stars of the SR variability type. The emission extends 
from --48 to --18 \kms, and therefore we confirm the large expansion velocity 
(\gsim\,12\,\kms) estimated by Olofsson et al. (1993a).

Each line profile was fitted with two parabolae in order to derive 
representative expansion velocities (Table~\ref{COfit}). 
We estimate the mass loss rates and photo-dissociation radii associated with
each component using the same approach as in Winters et al. (2003).
We adopt a CO/H mass ratio of 1$\times$10$^{-3}$. The differences in the 
mass loss rates and photo-dissociation radii estimated from the two lines 
are comparable to the uncertainties of the fits.

\begin{table}
\centering
\caption{CO line parameters of V1942\,Sgr. Formal uncertainties are given 
in parentheses.}
\begin{tabular}{cccccc}
\hline
\hline
         & \Vlsr & \Vexp & T$_{\rm mb}$ & \.M    & R$_{\rm CO}$  \\
         & \kms  & \kms  & K            & \Msold & 10$^{16}$ cm  \\
\hline
CO (1-0) & --33.75\,(0.25) & 17.5\,(0.5) & 0.12\,(0.01) & 6.1\,(0.3) 10$^{-7}$ & 6.9\,(0.2) \\
         & --33.75\,(0.25) &  5.0\,(0.5) & 0.10\,(0.01) & 1.0\,(0.1) 10$^{-7}$ & 4.0\,(0.2) \\
CO (2-1) & --33.25\,(0.25) & 17.75\,(0.5)& 0.25\,(0.02) & 4.3\,(0.3) 10$^{-7}$ & 5.7\,(0.2) \\
         & --33.25\,(0.25) &  5.0\,(0.5) & 0.20\,(0.02) & 6.9\,(0.5) 10$^{-8}$ & 3.2\,(0.2) \\
\hline
\end{tabular}
\label{COfit}
\end{table}

\section{V\,CrB}

V\,CrB is a metal-poor ([M/H] = --1.35) 
carbon star on the TP-AGB with a C/O ratio around 1.10 
(Abia et al. 2001). It is a Mira of period 358 days.
Using the period-luminosity relation for carbon Miras of Whitelock et al. 
(2006), Guandalini (2009) determines a luminosity of 5600 \Lsol ~and 
a distance of 547\,pc. \Teff ~is estimated at 2090 K (Bergeat et al. 2001). 
At such a low temperature (i.e. less than 2500 K) molecular hydrogen 
is expected to be the dominant species in the atmosphere and in the inner 
envelope of V CrB (Glassgold \& Huggins 1983). Indeed photospheric H$_2$ has 
been detected in the near-infrared (2.122 $\mu$m) by Johnson et al. (1983). 
This Mira is presently undergoing mass loss, since, for instance, it shows 
clear SiC dust emission at 11.3 $\mu$m (Goebel et al. 1981). 
The source has also been detected in the CO(1-0) and CO(2-1) 
rotational lines by Olofsson et al. (1993a), and more recently in CO(3-2) by 
Knapp et al. (1998). Contrary to V1942 Sgr, only one velocity component is 
visible. The central velocity is at \Vlsr = --99.0 \kms, 
the expansion velocity, \Vexp = 6.5 \kms, and the mass loss rate, 
\.M $\sim$ 2.1 10$^{-7}$ \Msold ~(at 547 pc). With the Plateau-de-Bure IRAM 
interferometer Neri et al. (1998) find a source of size of 7$''$ in CO (1-0). 
On the other hand IRAS has not detected extended far-infrared emission 
associated with V\,CrB (Young et al. 1993a, their Table 1). 

\begin{figure}
\centering
\includegraphics[width=8cm,angle=-90]{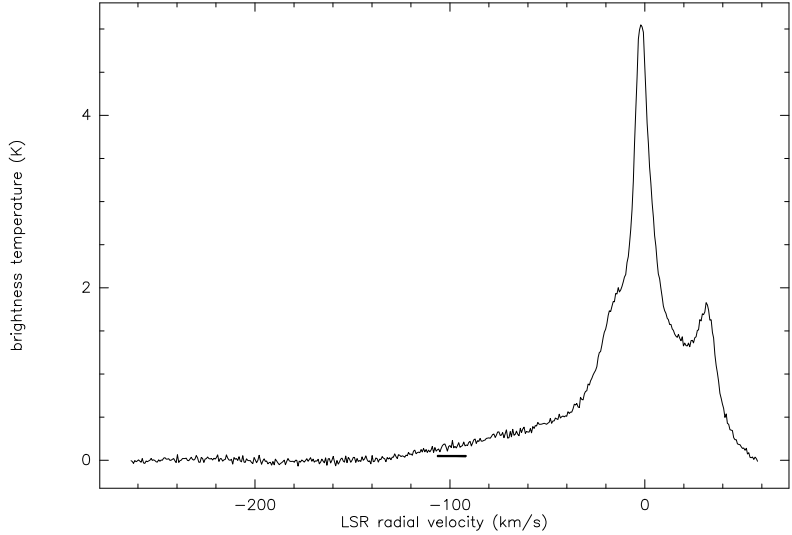}
\caption{Frequency-switch \HI 21 cm spectrum obtained with the NRT 
on the position of V\,CrB. The bar indicates the velocity range of the CO 
emission.}
\label{fswitchVCrB}
\end{figure}

V\,CrB was also observed in \HI with the NRT for a total of 44 hours.
The frequency-switched spectrum 
shows no feature around the expected velocity of 
--99.0 \kms ~(Fig.~\ref{fswitchVCrB}), and only 
a weak interstellar emission of at most 0.2\,K around --99\,\kms.

\begin{figure}
\centering
\includegraphics[width=10cm,angle=-90]{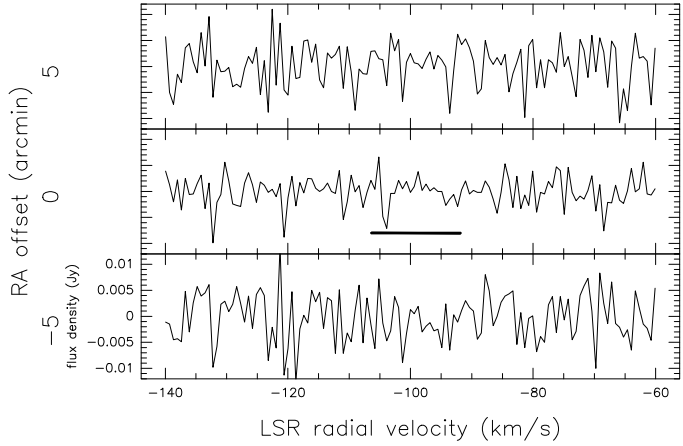}
\caption{\HI spectra obtained on V\,CrB (middle), and at +5$'$ east (top) 
and --5$'$ west (bottom) after correction for interstellar contamination 
(Sect.~3). The spectra have been smoothed to a resolution of 0.64 \kms. 
The bar indicates the velocity range of the CO emission. }
\label{HIVCrB}
\end{figure}

We obtained \HI data in the position-switch mode of observation with 
off-positions at $\pm$\,4$'$, $\pm$\,6$'$, $\pm$\,8$'$, $\pm$\,10$'$, 
$\pm$\,12$'$, $\pm$\,16$'$ $\pm$\,24$'$ and $\pm$\,32$'$.
As for V1942\,Sgr we find that the interstellar emission varies linearly 
with offset, in the velocity range --120 to --80 \kms. We are thus confident 
that the interstellar contamination can be corrected for accurately. 
The source is not detected at the star's position and at $\pm$\,5$'$ in RA
(Fig.~\ref{HIVCrB}, in which we have averaged the spectra obtained at 
+\,4$'$ and +\,6$'$, and at --\,4$'$ and --\,6$'$, in order to improve 
the sensitivity). By integrating over the velocity range defined by the 
CO emission, (i.e. --106 to --92 \kms), an upper limit of 
4 mJy$\times$\kms ~can be set on the intensity of the \HI emission at 
the V\,CrB's position in the area defined by the 4$'$$\times$22$'$ NRT beam. 
It translates to an upper limit of 3\,10$^{-4}$ \Msol ~in atomic hydrogen 
at 547 pc. As the source was not found to be extended by IRAS, we do not 
expect much material outside the NRT beam.

\section{Interpretation}

\subsection{V1942\,Sgr}

The CO line profiles observed in V1942 Sgr have a characteristic composite 
profile. This kind of profile has been interpreted as evidence for a 
succession of mass loss events with different outflow velocities by 
Knapp et al. (1998) and Winters et al. (2003). However the interferometric 
data obtained on EP Aqr, a source with such profiles, are difficult 
to explain with this scenario (Winters et al. 2007). Furthermore, 
in other cases, X Her (Kahane \& Jura 1996) and RS Cnc (Libert et al. 2009), 
there is evidence that the broad components originate in a bipolar flow. 
Bipolar flows are believed to develop at the end of the AGB phase when 
the stars are about to begin their evolution towards the planetary nebula 
phase (e.g. Sahai et al. 2007). 

The \HI spectrum obtained on the star's position shows a pedestal of width 
10 \kms ~that could be a counterpart of the narrow CO 
(1-0 and 2-1) components that have 
about the same width. The mass in hydrogen corresponding to this pedestal is 
$\sim$ 4 10$^{-3}$ \Msol. Assuming 90\,\% in H and 10\,\% in $^4$He, 
in number (i.e. a mean molecular weight, $\mu$, of 1.3), and adopting 
a mass loss rate of 1\,10$^{-7}$ \Msold ~from the narrow CO components 
(see Table 1), the timescale would be 6 10$^{4}$ years, and the 
radius 0.31 pc ($\equiv 2'$). The stellar effective temperature (2960 K) is 
large enough that we expect most of the hydrogen in atomic form.

On the other hand, there is no \HI counterpart to the broad CO components at 
a level of 2 mJy. This seems to indicate that the broad components correspond 
to a quite recent phenomenon. Indeed, adopting an upper limit of 2 mJy, the 
flux is $<$ 0.07 Jy$\times$\kms, and the mass in atomic hydrogen is at most 
5\,10$^{-3}$ \Msol. The timescale is then \lsim 10$^4$ years. 

The \HI spectra obtained on V1942\,Sgr are very similar to those obtained 
by Le\,Bertre \& G\'erard (2004) and Libert et al. (2007) on Y CVn, 
a well documented carbon star with 
a detached shell discovered by IRAS (Young et al. 1993a) and imaged by ISO 
(Izumiura et al. 1996). On the central position we detect a pedestal of width 
10 \kms, twice the expansion velocity measured for the narrow CO 
components. On all positions, we detect a narrow line 
of width, FWHM $\sim$ 3\,\kms. This narrow profile proves that the stellar 
wind from V1942\,Sgr is slowed down at some distance from the central star. 
Young et al. (1993b) have interpreted the detached shells that were revealed 
by IRAS at 60 $\mu$m as the effects of a slowing-down of stellar outflows by 
surrounding interstellar matter. Elaborating on this hypothesis and 
using the formalism of Lamers \& Cassinelli (1999), Libert et al. 
(2007) developed a model in which the inner radius of a detached shell 
corresponds to the location where the stellar wind is abruptly slowed down 
from \Vexp ~to $\sim$\,\Vexp/4. The outer radius corresponds to the location 
where external matter is compressed by the expanding shell (bow shock). 
They applied this model to Y CVn and obtained excellent fits to the \HI line 
profiles observed at different pointings, on and around the star's position. 

We are using the same model for V1942 Sgr. For the freely expanding wind 
(r$<$r$_1$) we adopt a velocity of 5.0 \kms, i.e. half the width of 
the pedestal, which corresponds also to the narrow CO components. 
The broad CO components have no obvious counterpart in H\,{\sc {i}}. 
They probably trace a short 
lived structure that is restricted to the central part of the circumstellar 
shell and that has no effect on the large scales probed at 21 cm. 
The mass loss rate, \.M = 1.0\,10$^{-7}$ \Msold, is adopted from the CO line 
fitting (Table 1). The central velocity is taken at --32.9 \kms.

\begin{figure}
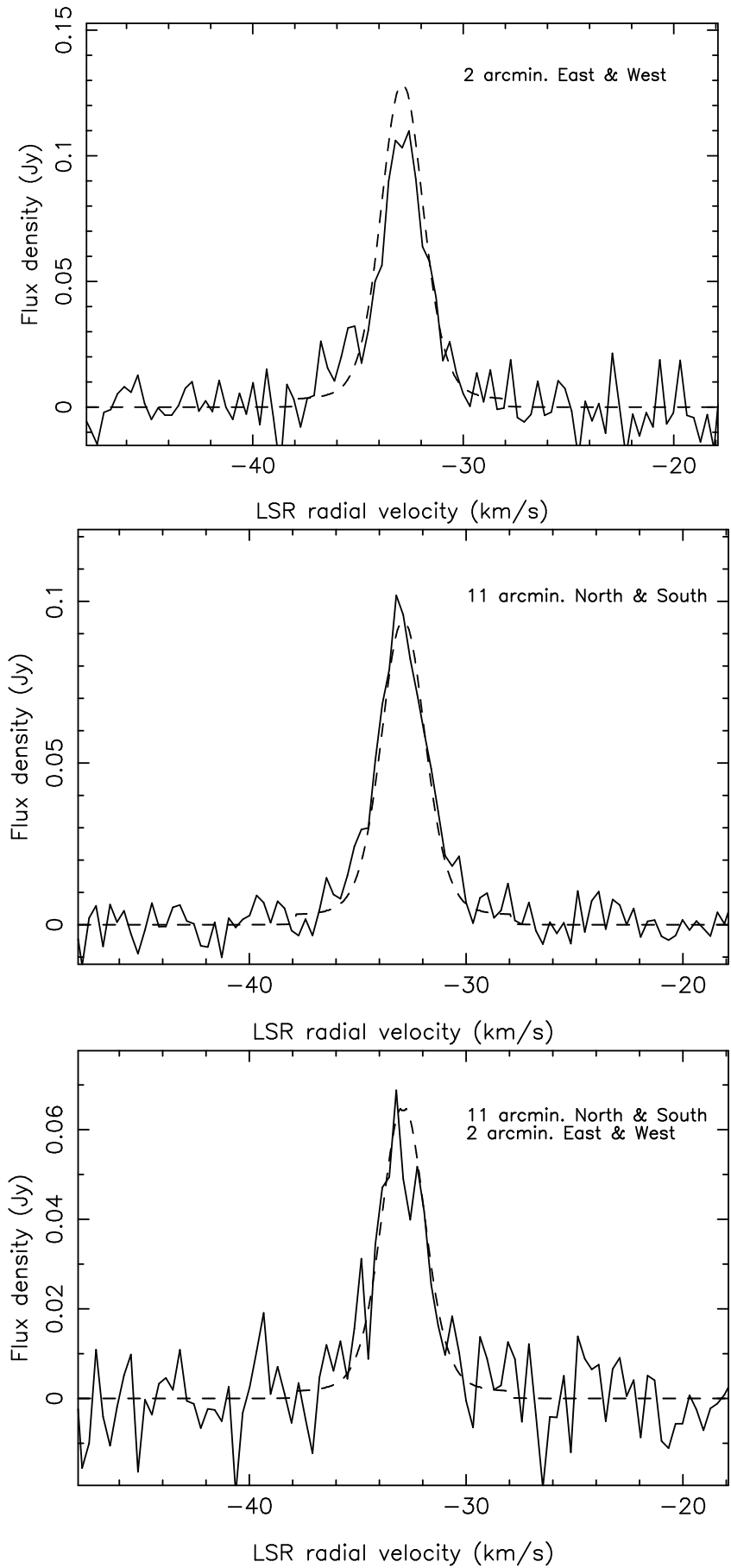

\centering
\includegraphics[width=8cm,angle=-90]{V1942-0p5EW.ps}
\includegraphics[width=8cm,angle=-90]{V1942-0p5NS.ps}
\includegraphics[width=8cm,angle=-90]{V1942-0p5NS-0p5EW.ps}
\caption{Comparison between the \HI line profiles obtained on V1942\,Sgr
and the detached-shell model discussed in Sect.~4.1. 
Top: average of the two spectra at +2$'$ (east) and --2$'$ (west).
Middle: average of the two spectra at +11$'$ (north) and --11$'$ (south).
Bottom: average of the four spectra at (+2$'$, +11$'$), (--2$'$, +11$'$), 
(+2$'$, --11$'$), and (--2$'$, --11$'$).
}
\label{compspectraV1942}
\end{figure}

We obtain a good fit to the different \HI line profiles 
(Figs.~\ref{centV1942} and \ref{compspectraV1942})
with the parameters given in Table~\ref{modelfit}. 
The external radius, r$_2 \sim 3'$, that we have adopted is compatible 
with that derived by Young et al. (1993a) from IRAS data at 60\,$\mu$m, 
r$_{\rm ext} \sim 3.2'$, but not the internal radius (r$_1 = 2'$, versus 
r$_{\rm int} \sim 0.2'$). In our model, r$_1$ is constrained by the parameters 
obtained from the low velocity CO components and by the intensity of the 
\HI pedestal. We assume that the inner shell is too small compared to 
the IRAS beam at 60 $\mu$m (FWHM\,$\sim 2'$) to have been reliably constrained.

\begin{table}
\centering
\caption{Model parameters (d = 535 pc).}
\begin{tabular}{ll}
\hline
\.M (in hydrogen)                 & 0.69 10$^{-7}$ \Msold\\
$\mu$                             & 1.3\\
t$_1$                             & 61\,10$^3$ years\\
t$_{DS}$                          & 5.4\,10$^5$ years\\
r$_1$                             & 0.31 pc (2.0$'$)\\
r$_f$                             & 0.41 pc (2.64$'$)\\
r$_2$                             & 0.47 pc (3.0$'$)\\
T$_0$($\equiv$ T$_1^-$), T$_1^+$  & 20 K, 746 K\\
T$_f$ (= T$_2$)                   & 81 K\\
v$_0$($\equiv$ v$_1^-$), v$_1^+$  & 5.0 \kms, 1.27 \kms\\
v$_f$                             & 0.066 \kms\\
v$_2$                             & 0.52 \kms\\
n$_1^-$, n$_1^+$                  & 0.45 H\,cm$^{-3}$, 1.8 H\,cm$^{-3}$\\
n$_f^-$, n$_f^+$                  & 21.0 H\,cm$^{-3}$, 2.15 H\,cm$^{-3}$\\
n$_2$                             & 1.66 H\,cm$^{-3}$\\
M$_{r < r_1}$ (in hydrogen)       & 4.2 10$^{-3}$ \Msol\\
M$_{DT,CS}$   (in hydrogen)       & 3.7 10$^{-2}$ \Msol\\
M$_{DT,EX}$   (in hydrogen)       & 6.3 10$^{-3}$ \Msol\\
\hline
\end{tabular}\\
\label{modelfit}

The notations are as in Libert et al. (2007). In particular, t$_{DS}$ 
is the formation time of the detached shell, M$_{DT,CS}$ is the mass of the 
circumstellar component of the detached shell, and M$_{DT,EX}$ is the external 
mass accreted in the detached shell.
\end{table}

\begin{figure}
\centering
\includegraphics[width=8cm,angle=-90]{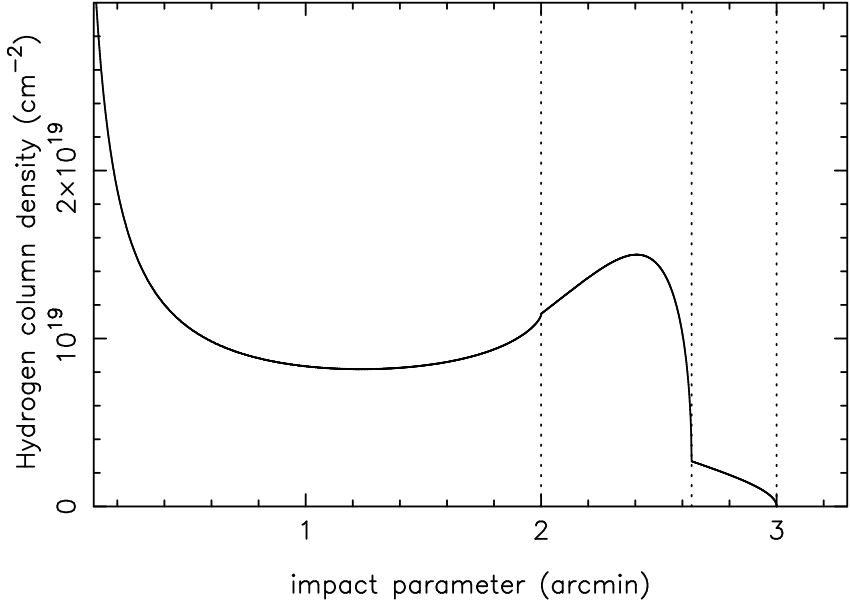}
\caption{Atomic hydrogen column density profile for the V1942\,Sgr model.
The vertical lines mark the radii r$_1$ (0.31 pc), r$_f$ (0.41 pc), and
r$_2$ (0.47 pc), which define the detached shell.}
\label{HIcolumndensity}
\end{figure}

\subsection{V\,CrB}

V CrB was not detected in H\,{\sc {i}}. As there is no significant Galactic 
confusion, we are quite confident on our upper limit of 3\,10$^{-4}$\,\Msol  
~in atomic hydrogen. For a source losing atomic hydrogen with a mass loss 
rate of 2.1\,10$^{-7}$\,\Msold, it would correspond to a timescale 
of 2000 years ($\mu$\,=\,1.3). However, 
the stellar effective temperature is so low (2090 K) that hydrogen should be 
in molecular form in the atmosphere and outwards (Glassgold \& Huggins 1983), 
until it is photo-dissociated by the interstellar radiation field. 
To estimate the distance, \Rph, at which  
this happens, we follow the approach of Morris \& Jura (1983). Assuming a mean 
intensity of the ultraviolet radiation between 912 and 1100 \AA ~of 
1.9\,10$^6$ photons\,cm$^{-2}$\,s$^{-1}$\,sr$^{-1}$, and that 0.11 of all the 
absorptions lead to a dissociation, we get \Rph\,$\sim$\,410\,\.M$^{1/2}$, 
with \Rph ~in pc and \.M in \Msold. For a mass loss rate of 
2.1\,10$^{-7}$\,\Msold, we obtain \Rph\,=\,0.2 pc ($\equiv$\,1.2$'$), which 
corresponds to a dynamical time of $\approx$\,3\,10$^4$\,years 
(\Vexp\,=\,6.5\,\kms). 

Therefore the non-detection of V CrB in \HI implies that it has not been 
undergoing mass loss at the present rate for more than 3.2\,10$^4$\,years. 
Furthermore, when comparing with V1942\,Sgr, which is at the same distance, 
we can state that V CrB has not gone through the same phase 
of mass loss during the past 5\,10$^5$\,years, because if it had done so 
it would have been easily detected like V1942\,Sgr.

This reasoning assumes that molecular hydrogen is not self-protected 
within small-scale structures that could develop in the stellar outflow. 
However the non-detection by IRAS of an extended emission around V CrB 
(Young et al. 1993a) agrees with our conclusion that mass loss has started  
only recently. 

\subsection{Discussion}

The V1942 Sgr proper motion measured by Hipparcos is 10.98 mas\,yr$^{-1}$ 
in RA and --5.10 mas\,yr$^{-1}$ in Dec. 
When corrected for solar motion towards apex, and for a distance of 
535\,pc, it translates to 6.45 mas in RA and --2.28 mas in Dec. This implies 
a motion in the plane of the sky at a velocity of 17 \kms, and ~at a position 
angle, PA = 110$^{\circ}$. Accounting for the radial velocity, 
\Vlsr = --33 \kms, we obtain a 3D space velocity of 37 \kms. 
The offset with respect to the central star that we find in 
the \HI map might therefore be an effect of the motion 
of V1942 Sgr relative to the surrounding ISM. Such a deformation in \HI has 
already been noted in several cases: Mira (Matthews et al. 2008), RX Lep 
(Libert et al. 2008),  RS Cnc (MR2007 and Libert et al. 2009). 
GL2006 noted also that many \HI sources are offset with respect to the 
central stars. A visual inspection of the IRAS map at 60\,$\mu$m 
of V1942 Sgr (Improved Reprocessing of the IRAS Survey\,: 
Miville-Desch\^enes
\& Lagache 2005) reveals that the image is slightly elongated in RA and 
shifted west by $\sim 1/2$ pixel ($\equiv 0.75'$), in agreement with our 
\HI map. Finally, it is worth noting that the central velocity in \HI 
is --32.9$\pm$0.3 \kms, whereas in CO it is --33.5$\pm$0.25 \kms. 
The effect is small 
but consistent with an interaction between the external shell of V1942 Sgr 
and its local ISM. Shifts in velocity of the \HI emission towards the LSR 
have already been reported in several red giants (GL2006, Matthews 
et al. 2008).

From their study of circumstellar shells resolved by IRAS, Young et al. (1993b)
find that, among nearby AGB stars detected in CO, Miras, in contrast to  
Semi-Regulars, are in general unresolved. They suggest that 
the latter have been losing matter for a longer time than the former. 
In their \HI survey of evolved stars GL2006  
obtained results that agree with this suggestion. Although their sample 
was small, several Miras with high mass loss rate could not be detected in 
H\,{\sc {i}}, whereas SRs were often easily detected. The high quality data 
that we have obtained on V1942\,Sgr and V\,CrB strengthen the case of SRs 
undergoing mass loss for a longer time than Miras. One normally assumes 
that SRs evolve into Miras, and it is puzzling to find no relics of this 
SR phase around several Miras. SRs might evolve directly in the post-AGB 
phase, as suggested also by the presence of bipolar outflows which has been 
reported in several cases (Kahane \& Jura 1996, Libert et al. 2009).

Young et al. (1993b) also find that extended sources are observed 
preferentially around carbon stars and GL2006 obtained a higher rate 
of detection of carbon stars in H\,{\sc {i}}.
However, the case of V CrB seems to suggest that some stars could reach 
the carbon-rich stage without undergoing substantial mass loss previously.
In a systematic investigation of the relations between mass loss 
and red giant characterstics, Winters et al. (2000) find 
that the mass loss rate depends critically on stellar parameters such as the 
effective temperature, which controls the dust formation, and the luminosity, 
which controls the radiation pressure. V CrB may have switched only 
recently from the B-regime (with a low and, presently, 
undetected wind) to the A-regime with a wind at a few 10$^{-7}$ \Msold.

Although both V1942\,Sgr and V\,CrB are carbon stars on the TP-AGB phase, 
their history of mass loss 
during the past 5\,10$^5$ years seem to differ radically. If it is correct 
that the bipolar shaping is a signpost of the end of the AGB, V1942 Sgr (and 
also the sources with composite CO line profiles) might be on the verge of 
leaving this phase. Both sources have about the same C/O abundance ratio,  
1.12 for V1942 Sgr (Olofsson et al. 1993b) and 1.10 for V\,CrB (Abia et al. 
2001), and the same luminosity, 5200 and 5600 \Lsol, respectively. 
Also both sources have a low  $^{12}$C/$^{13}$C abundance ratio,  
30 for V1942 Sgr (Abia \& Isern 1997) and 10 for V CrB (Abia et al. 2001), 
as compared to $\sim$ 40 for the majority of carbon stars in the AGB phase. 
The explanation of such low abundance ratios is not known, but could be 
due to a non-standard mixing process occurring in low-mass stars at the base 
of the convective stellar envelope (``cool bottom processing'', Nollett 
et al. 2003).

\section{Conclusions} 

The combination of high velocity resolution CO and \HI data  
is a promising tool to investigate the history of 
mass loss by evolved stars. The low level of Galactic \HI emission and 
the absence of small-scale structure in this emission have allowed us 
to obtain \HI data of high quality on V1942 Sgr and V CrB with the NRT. 
We have also obtained high quality CO (1-0) and (2-1) spectra of V1942 Sgr 
with the IRAM 30-m telescope.

For V1942 Sgr, the CO spectra exhibit composite profiles, that reveal 
a low velocity wind of $\sim$ 10$^{-7}$ \Msold ~and a high velocity wind, 
possibly bipolar, of $\sim$ 5\,10$^{-7}$ \Msold. The comparison with the \HI 
spectrum shows that this high velocity wind is recent with an age of at most 
10$^4$ years. On the other hand, 
the low velocity wind appears to have filled a cavity of 
$\sim$ 0.2 pc in radius and built the detached shell, that was discovered by 
IRAS, over a period of 5 10$^5$ years. Follow-up observations with the VLA 
and ALMA would help to improve this scenario, or possibly lead to 
a new scheme. The narrowness of the \HI 
line profile in V1942\,Sgr brings new evidence that AGB stellar winds are 
slowed down by their surrounding medium as surmised by Young et al. (1993b). 

For V CrB, the CO spectra that have been published reveal an outflow with 
expansion velocity, 6.5 \kms, and mass loss rate, 2.1 10$^{-7}$ \Msold. 
The non-detection in \HI of V CrB sets an upper limit of 3.2 10$^4$ years for
the age of this outflow. In such case of a star with low effective 
temperature, molecular hydrogen data are obviously needed to constrain 
better the history of mass loss.

\begin{acknowledgements}
The Nan\c{c}ay Radio Observatory is the Unit\'e scientifique de Nan\c{c}ay of 
the Observatoire de Paris, associated as Unit\'e de Service et de Recherche 
(USR) No. B704 to the French Centre National de la Recherche Scientifique 
(CNRS). The Nan\c{c}ay Observatory also gratefully acknowledges the financial 
support of the Conseil R\'egional de la R\'egion Centre in France. We thank 
the IRAM Director, P. Cox, for allowing the CO observations of V1942 Sgr to be 
made on Director's time (D01-09). 
IRAM is supported by INSU/CNRS (France), MPG (Germany), and IGN (Spain).
We are grateful to C. Abia, M. Busso, R. Guandalini, and A. Jorissen 
for enlightening discussions, and to the referee for careful comments 
that helped us to improve the manuscript.
This research has made use of the SIMBAD and VizieR databases, operated 
at CDS, Strasbourg, France and of the NASA's Astrophysics Data System.
\end{acknowledgements}

\end{document}